\documentclass{article}
\newcommand{\bfr}{\begin{flushright}}
\newcommand{\efr}{\end{flushright}}
 
\begin{document}
\title{Simple Models in Supersymmetric Quantum Mechanics on a 
Graph
}
\author{Nahomi Kan\footnote{kan@gifu-nct.ac.jp}
\\
{\small
Gifu National College of Technology,
Motosu-shi, Gifu 501-0495, Japan
}
\bigskip
\\
Koichiro Kobayashi\footnote{m004wa@yamaguchi-u.ac.jp}
\quad
and
\quad
Kiyoshi Shiraishi\footnote{shiraish@yamaguchi-u.ac.jp}\\
{\small
Yamaguchi University,
Yamaguchi-shi, Yamaguchi 753--8512, Japan}
}
\date{\today
}
\maketitle
\begin{abstract}
We study some sorts of dimensionally-deconstructed models for 
supersymmetric (Euclidean) quantum mechanics, or zero-dimensional field
theory.  In these models, we assign bosonic and fermionic
variables to vertices and edges of a graph.
We investigate a discrete version for the Gaussian model and the
Wess-Zumino-type
model on a graph. The topological index as a multiple integral
is discussed on these models.
In addition, we propose simple examples for supersymmetric extensions of
the Lee-Wick model and the Galileon model. A model with two
supersymmetries is also provided and generalization to `local'
supersymmtric models is examined.
\end{abstract}

\section{Introduction}
The non-perturbative effects play important roles in many aspects
of quantum field theory. Particularly, they are considered to be crucial
for breakdown of supersymmetry in field theories.
An approach to understand the dynamical supersymmetry breaking
is to study supersymmetric models in quantum mechanics, which has been
suggested by Witten \cite{WittenSQM}.
In a certain sense, quantum mechanics defined through the path integral
is equivalent to zero-dimensional field theory. Moreover, the view point
of the path integral provides prospects for topological properties of
dynamical models in many problems.
On the other hand, supersymmetric field theory has
topological nature of its own because of its cohomological structure
\cite{BBRT}. Therefore, various models with supersymmetry in
lower dimensions is worth studying 
due to mathematical interest. 

The present authors have examined field theories on a graph \cite{JMP}
and models with superfields on a graph \cite{PRD},
which are interpreted as 
extensions of dimensional deconstruction \cite{DD1,DD2}. The dimensional
deconstruction is a powerful tool to analyze the higher-dimensional
theory by adopting multiple fields in lower dimensions. Our models
mentioned above consist of different kinds of fields on vertices and
edges on a graph. Thus, we acquire an idea of `deconstructing' 
one-dimensional theory by assigning the different `multiplets' to
vertices and edges. The details are shown in Section~\ref{sec4}.

In the present paper, we propose various models for supersymmetric
quantum mechanics on a graph. In particular, we provide analogue models
for higher-derivative theories with supersymmetry.

Discrete models are also useful to make a functional integral well
defined mathematically. The models on a graph have a continuum
limit of discrete variables in simple restricted cases.
Actually, similar cases are known by
a lattice formulation of field theory in order to study
non-perturbative effects through numerical simulations
\cite{C1,C2,C3,K}. Another related study can be found in
recent literature on zero-dimensional models of matrix theory
\cite{KS1,KS2,KS3}. In the present paper, although we do not make
further mention on connection to these approaches and models, we
should keep our mind on possible development of our models by
incorporating their technical methods.
In general cases, our supersymmetric models on a graph do not have 
continuum limit. This feature is mathematically interesting and would be
studied in future. 

The organization of the present paper is as follows. In
Section~\ref{sec2}, we shall give a brief review of `zero-dimensional'
supersymmetry on a toy model. Section~\ref{sec3} is devoted to
description of an algebraic aspect of graph theory, namely,
introduction of matrices associated with a graph. After these
preparation, we construct a simple model on a graph in
Section~\ref{sec4}. An analogue model for a self-interacting field
theory is discussed in Section~\ref{sec5}. In Section~\ref{sec6}, we
deal with analogue models for higher-derivative theory. In
Section~\ref{sec7}, we consider a sort of extension, which leads to a
model with two supersymmetries. Based on this model, we explore the
possibility of `local' supersymmetry, requiring an individual parameter
of transformation for each variables in Section~\ref{sec8}. Finally, we
give concluding remarks in Section~\ref{sec9}.

\section{Review of zero-dimensional supersymmetry
\label{sec2}}
The simplest example \cite{BBRT} for supersymmetric quantum
mechanics contains bosonic (commuting) variables
$\phi$,
$F$ and fermionic (anticommuting, Grassmann) variables $\psi$,
$\bar{\psi}$.

The fermionic transformation is defined by using a supercharge $Q$ as
\begin{equation}
Q\phi=\psi\,,\quad Q\psi=0\,,\quad
Q\bar{\psi}=F\,,\quad QF=0\,.
\end{equation}
It is easy to find nilpotency $Q^2=0$ in these relations.
Then the following `action' $S$ is invariant under the transformation
induced by $Q$:
\begin{equation}
S[\phi,\psi,\bar{\psi},F]=F\cdot
P(\phi)-\bar{\psi}P'(\phi)\psi-\frac{1}{2}F^2\,,
\end{equation}
where $P$ is a function of $\phi$ and $P'$ is the first derivative of
$P$. One can find that the action $S$ is supersymmetric by construction,
because
$S$ can be written in the form
\begin{equation}
S=Q\left[\bar{\psi}\left(P(\phi)-\frac{1}{2}F\right)\right]\,.
\end{equation}

Here, we take $F$ as an auxiliary variable, so the action can be read
as 
\begin{equation}
S[\phi,\psi,\bar{\psi}]=\frac{1}{2}P^2(\phi)-\bar{\psi}P'(\phi)\psi\,.
\end{equation}
after elimination of the auxiliary variable $F$ by its equation of
motion. One can find this form of the action quite familiar, and
also find that
\begin{equation}
Z=\int_{-\infty}^\infty
\frac{d\phi}{\sqrt{2\pi}}\int
d\psi
d\bar{\psi}\,e^{-S[\phi,\psi,\bar{\psi}]}
\end{equation}
takes values $\pm 1$ or $0$ by case of the behavior of $P(\phi)$
in the limit of $\phi\rightarrow \pm\infty$. This is closely related to
the Witten index \cite{WittenIndex}.

In Section \ref{sec4} and subsequent sections, we utilize a number of 
variables and construct some models in supersymmetric quantum mechanics
on a graph; the method we adopt is very similar to the idea of
dimensional deconstruction \cite{DD1,DD2}.

\section{Review of graph theory and matrices therein
\label{sec3}}
In this section, we review some matrices which
are very useful to describe models on a graph \cite{JMP,PRD}.

Let $G({\cal V},{\cal E})$ be a graph with a vertex set ${\cal V}$ and
an edge set ${\cal E}$.
An oriented edge $e=[v,v']$ connects two adjacent vertices $v=o(e)$ and
$v'=t(e)$, where $o(e)$ is the origin of the edge $e$ and $t(e)$ is the
terminus of the edge $e$.
The number of adjacent vertices of a vertex $v$ is called the
degree of $v$, and is expressed by $d_v$.

The incidence matrix  $E$ for a directed graph is
defined by
\begin{equation}
E_{ve}(G)=\left\{
\begin{array}{cc}
1 & \mbox{if } v=o(e)\\
-1 & \mbox{if } v=t(e)\\
0 & \mbox{otherwise}
\end{array}
\right.\,.
\end{equation}

Similarly, the `unoriented' incidence matrix $B$ is
defined by
\begin{equation}
B_{ve}(G)=\left\{
\begin{array}{cc}
1 & \mbox{if } v=o(e)\mbox{~or~}v=t(e)\\
0 & \mbox{otherwise}
\end{array}
\right.\,.
\end{equation}

The graph Laplacian $L$ is defined by
\begin{equation}
L_{vv'}(G)=\left\{
\begin{array}{cc}
d_v & \mbox{if } v=v'\\
-1 & \mbox{if } [v, v']\in{\cal E}\\
0 & \mbox{otherwise}
\end{array}
\right.\,.
\end{equation}

There is an important relation between the incidence matrix and the
graph Laplacian for a graph,
\begin{equation}
L_{vv'}(G)=(E(G)E(G)^T)_{vv'}\,.
\end{equation}
Note that the Greek letter $\Delta$ is also used for representing
the graph Laplacian in many textbooks. We will not use the symbol in
this paper to avoid confusion with the difference operation.

For example, we consider a cycle graph.
A cycle graph $C_N$ is a set of $N$ ($N\ge 3$) vertices lined up along a
circle with edges between each vertex and its adjacent ones on each
side. The incidence matrix for a cycle graph $C_N$ is given
by
\begin{equation}
E(C_N)=\left(
\begin{array}{rrrrrr}
1 & 0 & 0 & \cdots& 0& -1\\
-1 & 1 & 0 &\cdots & 0&0\\
0 & -1 & 1 &\cdots &0&0\\
\vdots &\vdots &\vdots & \ddots&\vdots &\vdots\\
0 & 0 & 0 & \cdots & 1&0\\
0 & 0 & 0 & \cdots & -1&1
\end{array}
\right)\,.
\end{equation}
The cycle graph considered here is a closed circuit along with one
direction of edges, i.e., any vertex is an origin of one edge and a
terminus of another edge at the same time. The transposed matrix of the
incidence matrix can play a role of a difference operator.

The unoriented incidence matrix for a cycle graph $C_N$ is
\begin{equation}
B(C_N)=\left(
\begin{array}{rrrrrr}
1 & 0 & 0 & \cdots& 0& 1\\
1 & 1 & 0 &\cdots & 0&0\\
0 & 1 & 1 &\cdots &0&0\\
\vdots &\vdots &\vdots & \ddots&\vdots &\vdots\\
0 & 0 & 0 & \cdots & 1&0\\
0 & 0 & 0 & \cdots & 1&1
\end{array}
\right)\,.
\end{equation}

The graph Laplacian for $C_N$ is given by
\begin{equation}
L(C_N)=\left(
\begin{array}{rrrrrr}
2 & -1 & 0 & \cdots& 0& -1\\
-1 & 2 & -1 &\cdots & 0&0\\
0 & -1 & 2 &\cdots &0&0\\
\vdots &\vdots &\vdots & \ddots&\vdots &\vdots\\
0 & 0 & 0 & \cdots & 2&-1\\
-1 & 0 & 0 & \cdots & -1&2
\end{array}
\right)\,.
\end{equation}

One can find a similarity that
$\sum_{v'}L_{vv'}f_{v'}\approx -\partial_{t}^2 f(t)$,
where the parameter $t$ is discretized and is expressed
by a single sequence of $v$.
It is easy to compute the eigenvalue of $L(C_N)$, which is found to be
\begin{equation}
4\sin^2\frac{\pi p}{N}\qquad(p=0, 1, 2,\dots,N-1)\,.
\end{equation}
We can therefore see that the continuum limit
($N\rightarrow\infty$ and $\ell=Na=$constant) leads to $(2\pi/\ell)^2$
as the eigenvalue of the
$a^{-2}L(C_N)$, where $a$ is a distance scale or `lattice spacing',
namely $t$ is regarded as $na$ ($0\le n< N$). 

Another well-known graph is the path graph $P_N$, which possesses $N$
vertices and $N-1$ edges. The path graph has two ends (where $d_v=1$)
but all the other
$N-2$ vertices have degree two ($d_v=2$).
The incidence matrix for a path graph with a definite direction is
given by
\begin{equation}
E(P_N)=\left(
\begin{array}{rrrrrr}
1 & 0 & 0 & \cdots& 0\\
-1 & 1 & 0 &\cdots & 0\\
0 & -1 & 1 &\cdots &0\\
\vdots &\vdots &\vdots & \ddots&\vdots\\
0 & 0 & 0 & \cdots & 1\\
0 & 0 & 0 & \cdots & -1
\end{array}
\right)\,.
\end{equation}
The unoriented incidence matrix for a path graph $P_N$ is
similarly given as an $N\times (N-1)$ matrix, whereas
the graph Laplacian for $P_N$ is given by the following $N\times N$
matrix:
\begin{equation}
L(C_N)=\left(
\begin{array}{rrrrrr}
1 & -1 & 0 & \cdots& 0& 0\\
-1 & 2 & -1 &\cdots & 0&0\\
0 & -1 & 2 &\cdots &0&0\\
\vdots &\vdots &\vdots & \ddots&\vdots &\vdots\\
0 & 0 & 0 & \cdots & 2&-1\\
0 & 0 & 0 & \cdots & -1&1
\end{array}
\right)\,.
\end{equation}
(The graph Laplacian sometimes appears in textbooks of elementary
dynamics, to explain the vibration of a ball-spring system!)

\section{Construction of supersymmetric quantum mechanics on a
graph: Gaussian models
\label{sec4}}
Suppose $G({\cal V},{\cal E})$ be a simple graph.
We assign a scalar variable $\phi_v$ and a fermionic variable $\psi_v$
to each vertex $v$ of the graph, whereas a fermionic variable
$\bar{\psi}_e$ and a bosonic variable $F_e$ to each edge $e$ of
the graph.
Consider the following `action':
\begin{equation}
S[\phi,\psi,\bar{\psi},F]=\sum_{e\in{\cal
E}}\left[F_eP_{e}(\{\phi_v\})-\bar{\psi}_e\sum_{v\in{\cal
V}}\frac{\partial P_{e}(\{\phi_v\})}{\partial\phi_v}\psi_v-
\frac{1}{2}F_{e}F_{e}\right]\,,
\label{eq13}
\end{equation}
where $P_e$ is functions of $\phi_v$.

The action $S$ is invariant under the supersymmetry transformation
\begin{equation}
Q\phi_v=\psi_v\,,\quad
Q\psi_v=0\,,\quad
Q\bar{\psi}_e=F_e\,,\quad
QF_e=0\,.
\end{equation}
Note that $Q^2=0$ is ensured.
Note also that the supersymmetry transformation does not include
difference operators (matrices) in the present construction.

Because the action is given by
\begin{equation}
S=Q\left[\sum_{e\in {\cal
E}}\bar{\psi}_e\left(P_{e}(\{\phi_v\})-\frac{1}{2}F_e\right)\right]\,,
\end{equation}
the invariance under the fermionic transformation is trivial.

To define a Gaussian (free) model, we specify $P_e$ by a linear
combination of a few $\phi$'s:
\begin{equation}
P_e(\{\phi_v\})=\sum_{v\in {\cal V}}P_{ev}\phi_v\,,
\end{equation}
where $P_{ev}$ is a constant matrix.
The elimination of the auxiliary variables yields the action for a free
scalar and fermions:
\begin{equation}
S[\phi,\psi,\bar{\psi}]=\sum_{v,v'\in{\cal
V}}\sum_{e\in{\cal E}}\frac{1}{2}\phi_vP_{ve}^TP_{ev'}\phi_{v'}-
\sum_{v\in{\cal
V}}\sum_{e\in{\cal E}}\bar{\psi}_eP_{ev}\psi_v\,,
\label{8}
\end{equation}
where $\phi$ is a real scalar,
$\psi$ and $\bar{\psi}$ are fermions.

Here we consider the matrix as a combination of incidence matrices of a
graph:
\begin{equation}
P_{ev}=E^T+mB^T\,,
\end{equation}
where $E$ and $B$ are the oriented and unoriented incidence
matrices for a graph.
This is equivalent to the following expression:
\begin{equation}
P_e=\sum_{v\in{\cal V}}P_{ev}\phi_v=(1+m)\phi_{o(e)}-(1-m)\phi_{t(e)}\,.
\end{equation}

If we consider a cycle graph $C_N$, the eigenvalues of matrices can be
easily obtained and turns out to be%
\footnote{Similar procedure can be carried out for a graph
$G=C_{N_1}\cup C_{N_2}\cup
\cdots$.}
\begin{equation}
\det(E^T(C_N)+mB^T(C_N))=\prod_{p=0}^{N-1}[(1+m)-(1-m)e^{i\frac{2\pi
p}{N}}]\,,
\end{equation}
and
\begin{eqnarray}
&
&\det[(E(C_N)+mB(C_N))(E^T(C_N)+mB^T(C_N))]\nonumber \\
& &=\prod_{p=0}^{N-1}\left[4(1-m^2)\sin^2\frac{\pi
p}{N}+4m^2\right]\,.
\end{eqnarray}

Thus, in this case, the `partition function'
\begin{equation}
Z=\int [D\phi][D\psi][D\bar{\psi}]\,e^{-S[\phi,\psi,\bar{\psi}]}\,,
\end{equation}
where $[D\phi]=\prod_{v\in {\cal V}}\frac{d\phi_v}{\sqrt{2\pi}}$,
$[D\psi]=\prod_{v\in {\cal V}}d\psi_v$ and
$[D\psi]=\prod_{e\in {\cal E}}d\bar{\psi}_e$,
can be explicitly evaluated as a Gaussian multiple integral
and becomes
\begin{equation}
Z=\frac{\det(E^T(C_N)+mB^T(C_N))}
{\sqrt{\det[(E(C_N)+mB(C_N))(E^T(C_N)+mB^T(C_N))]}}=1\,.
\end{equation}
This answer is in agreement with that of Ref.~\cite{K}.

For a cycle graph, the transpose of the incidence matrix $E^T$
corresponds to a difference operator, which maps the difference
between variables assigned on adjacent vertices to the edge connecting
the vertices.
Of course, the `lattice spacing', which
is needed for correspondence to continuum theory, is considered as
being omitted here and can be recovered by rescaling variables
appropriately.

One can find that the transpose of the unoriented incidence matrix
provides the `mass term' or mass matrix in the action.
It is worth noting that the mass term is slightly `non-local', due to
its concerning with the variables on the nearest neighbor vertices (or
edges). This is however a natural choice because of the assignment of
variables to make the supersymmetry apparent.

\section{Superpotentials on a cycle graph
\label{sec5}}
Next, we will examine non-linear models.
For simplicity, in this section, we mainly consider a cycle graph $C_N$
and generalization to other graphs will be briefly discussed later.
Thus, we simply denote the incidence matrix as $E$.

Consider the case that $P_e$ in (\ref{eq13}) is a sum of the
`difference'
$\sum_{v\in{\cal V}}E^T_{ev}\phi_v$ and non-linear functions
${\it\Delta}W_e(\{\phi_v\})$. In Ref.~\cite{IJMPA}, we considered
similar discrete models in which a kink-shaped configuration exists.
This is equivalent to choosing the function $P_e$ in the present paper
as 
\begin{eqnarray}
P_e&=&\sum_{v\in{\cal
V}}E^T_{ev}\phi_v+{\it\Delta}W_e(\{\phi_v\})\nonumber
\\
&=&\phi_{o(e)}-\phi_{t(e)}+g\left(1-\frac{\phi_{o(e)}^2+\phi_{o(e)}
\phi_{t(e)}+\phi_{t(e)}^2}{3}\right)\,,
\end{eqnarray}
up to coefficients. Here $g$ is a coupling constant.

The advantage of this choice is the fact that the bosonic action
reduces to the simple sum of the `kinetic' term and the `potential'
term. That is,
\begin{equation}
\sum_{e\in{\cal
E}}(P_e)^2=\sum_{e\in{\cal
E}}(\phi_{o(e)}-\phi_{t(e)})^2+g^2\sum_{e\in{\cal
E}}\left(1-\frac{\phi_{o(e)}^2+\phi_{o(e)}
\phi_{t(e)}+\phi_{t(e)}^2}{3}\right)^2\,,
\end{equation}
because
\begin{eqnarray}
& &\sum_{e\in{\cal
E}}(\phi_{o(e)}-\phi_{t(e)})\left(1-\frac{\phi_{o(e)}^2+\phi_{o(e)}
\phi_{t(e)}+\phi_{t(e)}^2}{3}\right)\nonumber\\
& &=\sum_{e\in{\cal
E}}\left[\left(\phi_{o(e)}-\frac{\phi_{o(e)}^3}{3}\right)-
\left(\phi_{t(e)}-\frac{\phi_{t(e)}^3}{3}\right)\right]=0\,.
\end{eqnarray}
For the cycle graph, in which every vertex has degree two,
the cancellation of the cross term is obvious. Actually,
the cancellation is attained if  
${\it\Delta}W_e(\phi_{o(e)},\phi_{t(e)})=
{\it\Delta}W_e(\phi_{t(e)},\phi_{o(e)})$ for a cycle graph,
owing to its homogeneous structure. In the above example, however,
cancellation occurs only in the terms corresponding to the
nearest-neighbor edges. We can imagine such a `locality' in a model if
${\it\Delta}W_e$ is expressed by
\begin{equation}
{\it\Delta}W_e\equiv
\frac{W(\phi_{o(e)})-W(\phi_{t(e)})}{\phi_{o(e)}-\phi_{t(e)}}\,.
\end{equation}
Note that for $W(\phi)=m\phi^2$, we get
${\it\Delta}W_e=m\sum_{v\in{\cal V}}B^T_{ev}\phi_v$, which has appeared
in the Gaussian model in the previous section.

Now, we examine the topological index, the partition function of the
model.
First, we consider a parametrized partition function
\begin{equation}
Z(t)=\int[D\phi][D\psi][D\bar{\psi}][DF]
e^{-S[\phi,\psi,\bar{\psi},F]+tQV}\,,
\end{equation}
which turns out to be independent of the parameter $t$, so $Z(t)=Z$, if
$QS=0$ and
$Q^2=0$.
In the expression of $Z(t)$, $F_e$ should be replaced as
$F_e\rightarrow iF_e$ for convergence of integration, so the action
becomes
\begin{equation}
S[\phi,\psi,\bar{\psi},F]=\sum_{e\in{\cal
E}}\left[iF_eP_{e}(\{\phi_v\})-\bar{\psi}_e\sum_{v\in{\cal
V}}\frac{\partial P_{e}(\{\phi_v\})}{\partial\phi_v}\psi_v+
\frac{1}{2}F_{e}F_{e}\right]\,.
\end{equation}
Now, we take $tQV=\frac{1}{2}\sum_{e\in{\cal E}}F_e^2$. After carrying
out fermionic integrations, we obtain
\begin{eqnarray}
Z&=&\int[D\phi][DF]\left|\frac{\partial
P_{e}(\{\phi_v\})}{\partial\phi_v}\right|
\exp\left[-i \sum_{e\in{\cal E}}F_eP_e\right]\nonumber \\
&=&\int_{\Omega_P}[DP]\,\delta^N(P)
\,.
\end{eqnarray}
Here we denote $\int_{\Omega_P}[DP]=\int[D\phi]\left|\frac{\partial
P_{e}(\{\phi_v\})}{\partial\phi_v}\right|$, where the determinant is
the Jacobian. Note that the integration region $\Omega_P$ does not need
to be
$R^N$. Therefore,
$Z$ gives the winding number of map
$P_e(\{\phi_v\})$.%
\footnote{The manifold represented by the potential in the continuum
limit is usually taken to be connected, so the winding number
($R^N\rightarrow \Omega_P$, where
$R^N\cup\{\infty\}\approx\Omega_P\cup\{\infty\}\approx S^N$) should be
$0,\pm 1$.} We find that, under the assumption of `locality', for
$W(\phi)\propto
\phi^n$ ($n$ is an integer),
$Z=0$ if
$n$ is odd while 
$Z=1$ (or $-1$) if $n$ is even.
This is the same as the result on usual supersymmetric quantum
mechanics.
It is noteworthy that the analysis of the partition function
is easy if ${\it\Delta}W_e$ is a function only of $\phi_{o(e)}$ and
$\phi_{t(e)}$.

Before closing this section, we give a comment on generalization to the
model for quantum mechanics on a general graph.
The separation of the kinetic term and the potential term is possible
if we elaborate to cancel the cross term in $\sum_{e\in{\cal E}} P_e^2$.
To accomplish the `local' cancellation as in the case with cycle
graphs, we should choose an Euler graph. For every vertex $v$ in an
Euler graph, the number of edges satisfying $o(e)=v$ equals to the
number of edges satisfying $t(e)=v$. The cancellation thus occurs at
every vertex.

\section{`Higher-derivative' models
\label{sec6}}
Recently, higher-derivative models in field theory have been eagerly
studied in particle physics \cite{GOW} and cosmology \cite{DGP,NRT,CF}.
Their supersymmetric generalizations are also investigated by many
authors \cite{ADG1,ADG2,DPSS,GSS,KLO}.
In this section, we consider quantum-mechanical analogue models of
supersymmetric higher-derivative theories.

Consider $K_{ee'}(\{\phi_v\})$,  a matrix of functions on $\phi_v$ and
assume that $K_{ee'}$ is a symmetric matrix,
i.e., $K_{ee'}=K_{e'e}$.

Now, the action for the supersymmetric model including $K_{ee'}$ becomes
\begin{eqnarray}
S&=&Q\left[\sum_{e,e'\in{\cal
E}}\bar{\psi}_eK_{ee'}(\{\phi_v\})\left(P_{e'}(\{\phi_v\})-
\frac{1}{2}F_{e'}
\right)\right]\nonumber \\
&=&\sum_{e,e'\in{\cal
E}}\left[F_eK_{ee'}P_{e'}-\bar{\psi}_e\sum_{v\in{\cal V}}\frac{\partial
(K_{ee'}P_{e'})}{\partial\phi_v}\psi_v+
\frac{1}{2}\bar{\psi}_{e}\sum_{v\in{\cal V}}\frac{\partial
K_{ee'}}{\partial\phi_v}\psi_vF_{e'}\right.\nonumber \\
& &\qquad\quad\left.-
\frac{1}{2}F_{e}K_{ee'}F_{e'}\right]\,.
\end{eqnarray}

The equation of motion for $F_e$ turns out to be
\begin{equation}
\sum_{e'\in{\cal E}}K_{ee'}P_{e'}+
\sum_{e'\in{\cal E}}\sum_{v\in{\cal
V}}\frac{1}{2}\bar{\psi}_{e'}\frac{\partial
K_{e'e}}{\partial\phi_v}\psi_v-
\sum_{e'\in{\cal E}}K_{ee'}F_{e'}=0\,,
\end{equation}
and reduces to
\begin{equation}
F_e=P_{e}+
\sum_{e',e''\in{\cal E}}\sum_{v\in{\cal
V}}\frac{1}{2}K_{ee''}^{-1}\frac{\partial
K_{e''e'}}{\partial\phi_v}\bar{\psi}_{e'}\psi_v\,.
\end{equation}
Substitution of the equation simplifies the action to
\begin{eqnarray}
S&=&\sum_{e,e'\in{\cal
E}}\left[\frac{1}{2}P_{e}K_{ee'}P_{e'}-\bar{\psi}_e\sum_{v\in{\cal
V}}K_{ee'}\frac{\partial P_{e'}}{\partial\phi_v}\psi_v\right.\nonumber
\\ & &+\left.\frac{1}{8}\sum_{e'',e'''\in{\cal E}}\sum_{v,v'\in{\cal
V}}\bar{\psi}_{e}\psi_{v}\frac{\partial
K_{ee''}}{\partial\phi_{v}}K_{e''e'''}^{-1}\frac{\partial
K_{e'''e'}}{\partial\phi_{v'}}\bar{\psi}_{e'}\psi_{v'}\right]\,.
\end{eqnarray}

If we incorporate the incidence matrix $E$ and its transpose $E^T$ as
the `difference operators' into $K_{ee'}$, we can construct discrete
analogue model for higher-derivative theories.

The partition function can be evaluated as in a similar manner shown in
the previous section. In the present
case, we choose $tQV=\frac{1}{2}\sum_{e\in{\cal E}}F_e(KP)_e$. 
(Here and hereafter the sum over edges (or vertices) is not
indicated if the multiplication of matrices is obvious.) Thus, we
get
\begin{eqnarray}
Z&=&\int[D\phi][DF]\left|\frac{\partial
(K_{}P)_{e}}{\partial\phi_v}\right|
\exp\left[-i \sum_{e\in{\cal E}}F_e(K_{}P)_e\right]\nonumber \\
&=&\int_{\Omega_{KP}}[D(KP)]\,\delta^N(KP)
\,,
\label{KP}
\end{eqnarray}
where the inner sum over edges is suppressed.

\subsection{Lee-Wick model}
More than forty years ago, Lee and Wick and the other authors have
considered higher-derivative action in order to avoid the infinity in
quantum field theory \cite{LW1,LW2,LW3,LW4}.
Recently, the idea has been revived for solving the hierarchy problem
\cite{GOW}. In this subsection, we provide a discrete model for quantum
mechanics with `higher derivatives'.

We adopt the following matrix as $K_{ee'}$:
\begin{equation}
K_{ee'}=\delta_{ee'}+\alpha(E^TE)_{ee'}\,,
\end{equation}
where $\delta_{ee'}$ denotes the identity matrix and $\alpha$ is a
constant.  Moreover, we
consider the simplest case, therefore we take
\begin{equation}
P_{e'}=\sum_{v\in{\cal V}}E^T_{e'v}\phi_v\,.
\end{equation}
The supersymmetric action then becomes
\begin{eqnarray}
S[\phi,\psi,\bar{\psi},F]&=&\sum_{e\in{\cal
E},v\in{\cal V}}F_e(E^T+\alpha E^TEE^T)_{ev}\phi_v\nonumber \\
& &-\sum_{e\in{\cal
E},v\in{\cal V}}\bar{\psi}_e(E^T+\alpha E^TEE^T)_{ev}\psi_v\nonumber \\
& &-
\sum_{e,e'\in{\cal
E}}\frac{1}{2}F_{e}(\delta_{ee'}+\alpha E^TE)_{ee'}F_{e'}\,.
\end{eqnarray}
Eliminating the auxiliary fields $F_e$, we are left with
\begin{eqnarray}
S[\phi,\psi,\bar{\psi}]&=&\frac{1}{2}\sum_{v',v\in{\cal
V}}\phi_{v'}(EE^T+\alpha EE^TEE^T)_{v'v}\phi_v\nonumber \\ &
&-\sum_{e\in{\cal E},v\in{\cal
V}}\bar{\psi}_e(E^T+\alpha E^TEE^T)_{ev}\psi_v\,.
\end{eqnarray}
This model is an analogue of Lee-Wick theory, whose Lagrangian is
${\cal L}=-\frac{1}{2}\phi\nabla^2(1+M^{-2}\nabla^2)
\phi-\bar{\psi}D(1+M^{-2}\nabla^2)\psi$,
where $\nabla^2$ is the Laplacian and $D$ denotes the Dirac operator
\cite{ADG1,ADG2,DPSS,GSS}, with $M^{-2}\sim\alpha a^2$ ($a$ is a
length scale).

The topological value for the partition function of this model turns
out to be unity for ${\it\Delta}W_e\ne 0$, in general.  
The variable $(KP)_e$ in (\ref{KP}) depends not only on ${\it\Delta}W_e$
but also on
${\it\Delta}W_{\tilde{e}}$, where $\tilde{e}$ is the edge whose end is
the same with one of the edge $e$. Thus,  even if some
${\it\Delta}W_e$ takes a restricted value bound above or below, 
the value for $(KP)_e$ can run over from $-\infty$ to $\infty$, in
general. 


\subsection{Galileon model}
In cosmology, scalar field theories with higher-derivative terms are
studied with much interest.
The DGP-like Galileon term, which is cubic in a scalar field with four
derivative operators, has been motivated from D-brane theory
\cite{DGP}.
The generalized Galileon field theory has been developed recently
\cite{NRT,CF}. Furthermore,
an attempt to supersymmetrize the Galileon models appears in
Ref.~\cite{KLO}.

Here we choose $K_{ee'}$ for an analogue model:
\begin{equation}
K_{ee'}=\delta_{ee'}+\beta_1\sum_{v\in{\cal V}}E^T_{ev}\phi_vE_{ve'}+
\frac{\beta_2}{4}[(E^TE)_{ee'}(B\phi)_{e'}+(B\phi)_{e}(E^TE)_{ee'}]\,,
\end{equation}
where the coefficients $\beta_1$  and $\beta_2$ are  constant.
Considering now the simplest case:
\begin{equation}
P_{e'}=\sum_{v\in{\cal V}}E^T_{e'v}\phi_v\,,
\end{equation}
with help of the following identity,
which is trivial if one rewrite this using $v=o(e)$ and $v=t(e)$:
\begin{equation}
2\sum_{v\in{\cal V}}E^T_{ev}f_vg_v=\sum_{v\in{\cal V}}E^T_{ev}f_v
\sum_{v'\in{\cal V}}B^T_{ev'}g_{v'}+\sum_{v\in{\cal V}}E^T_{ev}g_v
\sum_{v'\in{\cal V}}B^T_{ev'}f_{v'}\,,
\label{id}
\end{equation}
we obtain the part of the action for scalars:
\begin{eqnarray}
S_B[\phi]&=&\frac{1}{2}\sum_{e,e'\in{\cal E}}P_eK_{ee'}P_{e'}\nonumber
\\ &=&\frac{1}{2}\sum_{v,v'\in{\cal
V}}\phi_v(EE^T)_{vv'}\phi_{v'}+\frac{\beta_1}{2}
\!\!\!\!\!\!\!
\sum_{v,v',v''\in{\cal
V}}
\phi_v(EE^T)_{vv'}\phi_{v'}(EE^T)_{v'v''}\phi_{v''}\nonumber
\\ 
& &+\frac{\beta_2}{4}\sum_{v,v'\in{\cal
V}}
\phi_v(EE^TEE^T)_{vv'}\phi_{v'}\phi_{v'}\,.
\end{eqnarray}

The bilinear operator 
\begin{equation}
\Gamma(f,g)_v=\frac{1}{2}\sum_{v'\in{\cal V}}\{L_{vv'}f_{v'}g_{v'}-
f_vL_{vv'}g_{v'}-g_vL_{vv'}f_{v'}\}\,
\end{equation}
has been introduced by Chung, Lin and Yau recently \cite{LY,CLY,BHLLMY}
(but they used the other type of Laplacian as $L$).
The correspondence to continuum theory is known as
$\Gamma(f,g)_v\sim -a^2\partial f\cdot\partial g$ (where $a$ is a
lattice spacing). Therefore, if we choose $\beta_2=-\beta_1=\beta$, the
bosonic part of the action can be read as
\begin{eqnarray}
S_B[\phi]&=&\frac{1}{2}\sum_{v\in{\cal
V}}(E^T\phi)_v(E^T\phi)_{v}+\frac{1}{2}\beta
\sum_{v\in{\cal
V}}
(EE^T\phi)_{v}\Gamma(\phi,\phi)_v\,,
\end{eqnarray}
and the action for continuum theory derived from this can be written as
$S\sim \int dt
[\frac{1}{2}(\partial\phi)^2+\frac{\beta a^2}{2}\partial^2\phi
(\partial\phi)^2]$, which is the action for DGP-type Galileon.

If we regard the incidence matrice as a difference operator, a
continuum limit can be achieved up to some distance scale $a$  and we
get
\begin{equation}
K_{ee'}\Rightarrow
K=1+\beta_1a^2\stackrel{\leftarrow}{\partial}\phi
\stackrel{\rightarrow}{\partial}+
\frac{\beta_2a^2}{2}[\stackrel{\leftarrow}{\partial}\stackrel{\rightarrow}{\partial}\phi+\phi
\stackrel{\leftarrow}{\partial}\stackrel{\rightarrow}{\partial}]\,,
\end{equation}
where the arrows ($\rightarrow$, $\leftarrow$) indicate the direction
which the derivative operator  acts on. It will be interesting to
examine the system governed by the action
$S=\int dt \frac{1}{2}P(\phi)K(\phi)P(\phi)$ with
$P(\phi)=\stackrel{\rightarrow}{\partial}\phi+W'(\phi)$, where
$W'(\phi)$ is a certain function of $\phi$. We also imagine
generalization to higher-dimensional scalar models. In such a manner,
the quantum mechanical model with rigid supersymmetry provides a new
insight to model building in field theory. In any case, because
investigation of continuum models is beyond the scope of the present
paper, these subjects are left for future study.

\section{Models with two supersymmetries
\label{sec7}}
The models so far considered is constructed by variables assigned to
vertices and edges of a graph. In general graphs, the numbers of
vertices and edges are different, whereas they coincides with each other
for cycle graphs. Therefore, there are `zero modes' of the matrices
associated with general graphs a priori. As a zero-dimensional model,
the partition function becomes trivial in such a case without discarding
zero-mode contributions.

In this section, we improve the assignment of the bosonic and fermionic
variables; both variables are assigned vertices as well as edges.
This extension enables us to consider two supersymmetries in a model.
This formulation is useful to attempt to consider `local
supersymmetry' in the next section, but we will find difficulty in the
establishment.

Suppose that scalar variables are assigned to each edge as well as
to each vertex, that is
\begin{equation}
\phi=\left(
\begin{array}{c}
\phi_v \\
\phi_e 
\end{array}\right)\,,
\end{equation}
and similarly $\psi$, $\bar{\psi}$ and $F$ are put on both vertices and
edges of a graph:
\begin{equation}
\psi=\left(
\begin{array}{c}
\psi_v \\
\psi_e 
\end{array}\right)\,,\quad
\bar{\psi}=\left(
\begin{array}{c}
\bar{\psi}_v \\
\bar{\psi}_e 
\end{array}\right)\,,\quad
F=\left(
\begin{array}{c}
F_v \\
F_e 
\end{array}\right)\,.
\end{equation}

Now we introduce two fermionic transformation induced by supercharges
$Q_1$ and
$Q_2$. They are expressed as
\begin{equation}
Q_1\phi=\psi\,,\quad Q_1\psi=0\,,\quad Q_1\bar{\psi}=F\,,\quad Q_1F=0\,,
\end{equation}
and
\begin{equation}
Q_2\phi=\bar{\psi}\,,\quad Q_2\psi=-F\,,\quad Q_2\bar{\psi}=0\,,\quad
Q_2F=0\,.
\end{equation}
One can see that the supersymmetry algebra takes the form
\begin{equation}
Q_1^2=Q_2^2=0\,,\qquad
Q_1Q_2+Q_2Q_1=0\,.
\end{equation}

To construct an invariant `action' of the variables,
we define the following matrix for convenience:
\begin{equation}
{\bf E}=\left(
\begin{array}{cc}
O & -E\\
E^T & O
\end{array}\right)\,.
\end{equation}
Note that this square matrix satisfies ${\bf E}^T=-{\bf E}$.
The following action is invariant under the two transformations:
\begin{eqnarray}
S[\phi,\psi,\bar{\psi},F]&=&Q_1\left[\bar{\psi}^T\left({\bf
E}\phi-\frac{1}{2}F\right)\right]=Q_2\left[{\psi}^T\left({\bf
E}\phi+\frac{1}{2}F\right)\right]\nonumber \\ &=&F^T{\bf
E}\phi-\bar{\psi}^T{\bf E}\psi-\frac{1}{2}F^TF\,.
\end{eqnarray}
Eliminating the auxiliary fields $F$, we obtain
\begin{equation}
S[\phi,\psi,\bar{\psi}]=\frac{1}{2}({\bf
E}\phi)^T{\bf
E}\phi-\bar{\psi}^T{\bf E}\psi\,.
\end{equation}
This action has a very similar form to the action for free fields.

Conversely, we can suppose a simpler system by identifying the two
fermionic  species as $\psi=\bar{\psi}=\Psi$. Then the action
\begin{equation}
S_0[\phi,\Psi]=\frac{1}{2}({\bf
E}\phi)^T{\bf
E}\phi-\frac{1}{2}{\Psi}^T{\bf E}\Psi\,.
\label{2s}
\end{equation}
is invariant under a fermionic transformation induced by a supercharge
$Q$, i.e.,
$QS_0[\phi,\Psi]=0$, provided that the transformation rules are given by
\begin{equation}
Q\phi=\Psi\,,\qquad Q\Psi=({\bf E}\phi)\,.
\label{gt1}
\end{equation}
Note that $Q^2={\bf E}$.

The simple action (\ref{2s}) is useful to investigate the possibility
of introducing further coupling to other variables.
We will examine such a case via considering `local' supersymmetry in the
next section.

\section{A `locally' supersymmetric model?
\label{sec8}}
A simple locally supersymmetric model has been offered by van
Nieuwenhuizen \cite{vN1,vN2,vN3}. 
His model contains a massless free scalar field and a
massless free fermionic field. We now come to an idea of constructing a `locally' supersymmetric model
which has the same structure as a simple model, by extending the last
model.
Therefore, we consider the last
model and the `local' definition of supersymmetry.

\subsection{difficulty in `local' models}
We first restrict ourselves on a model on a cycle graph $C_N$ here.
We start with defining the `local' super-transformation:
\begin{equation}
\delta\phi_c=\epsilon_c\Psi_c\,,\quad
\delta\Psi_c=({\bf
E}\phi)_c(\tilde{\epsilon})_c\,,
\label{lt}
\end{equation}
where $c=1,\dots,2N$, $\tilde{\epsilon}\equiv\frac{1}{2}{\bf
B}\epsilon$, with
\begin{equation}
{\bf B}=\left(
\begin{array}{cc}
O & B(C_N)\\
B(C_N)^T & O
\end{array}\right)\,.
\end{equation}
The `locality' we consider here should be the property
that the transformation on a variable includes a limited number of
variables in the neighbor vertices and edges. 

Then the variation of the action (\ref{2s}) becomes
\begin{eqnarray}
\delta S_0
&=&({\bf
E}\phi)^T{\bf
E}\delta\phi-{\Psi}^T{\bf E}\delta\Psi\nonumber \\
&=&\sum_c\left[({\bf
E}\phi)_c\sum_a{\bf
E}_{ca}\epsilon_a\Psi_a+({\bf E}\Psi)_c({\bf
E}\phi)_c(\tilde{\epsilon})_c\right]\nonumber\\
&=&\sum_c\left[({\bf E}\epsilon)_c({\bf
E}\phi)_c(\tilde{\Psi})_c\right]\,,
\label{eqw}
\end{eqnarray}
where $\tilde{\Psi}\equiv\frac{1}{2}{\bf B}\Psi$.
Up to now, the definition of the `local' transformation (\ref{lt}) seems
to be good, nevertheless the transformation is slightly non-local
because of the use of $\tilde{\epsilon}$. 
The identity
(\ref{id}) has been used in the second line of Eq.~(\ref{eqw}) and it
enables us to write down the result as a single term.

To compensate this variation, we introduced a new
Grassmann variable
$\lambda_c$, which has its own variation
\begin{equation}
\delta\lambda_c=({\bf E}\epsilon)_c+\cdots\,,
\end{equation}
and we add a term into the action:
\begin{equation}
S_N=-\sum_c\left[\lambda_c({\bf
E}\phi)_c(\tilde{\Psi})_c\right]\,.
\end{equation}
This is the very orthodox way to obtain local symmetries in
field theory.
The actual variation of the additional term can be found as:
\begin{eqnarray}
\delta S_N&=&-\sum_c\left[({\bf E}\epsilon)_c({\bf
E}\phi)_c(\tilde{\Psi})_c\right]\nonumber \\
&-&\sum_c\left[\lambda_c\sum_a{\bf
E}_{ca}\epsilon_a\Psi_a(\tilde{\Psi})_c\right]\nonumber \\
&-&\sum_c\left[\lambda_c({\bf
E}\phi)_c\sum_a\frac{1}{2}{\bf B}_{ca}({\bf
E}\phi)_a(\tilde{\epsilon})_a\right]\,.
\label{vn}
\end{eqnarray}
The first term of course cancels $\delta S_0$ given above.
The second term in Eq.~(\ref{vn}) can be removed if we consider the
additional variation in $\delta\Psi_c$ as
\begin{equation}
\delta\Psi_c=({\bf
E}\phi)_c(\tilde{\epsilon})_c-\lambda_c(\widetilde{\Psi\epsilon})_c\,.
\end{equation}
To see this, we use the identity (\ref{id})
and $\lambda_c\lambda_c=\Psi_c\Psi_c=0$.

If we attempt to cancel the third term,
we need a new variable $h_{ab}$, which has the variation 
\begin{equation}
\delta h_{ab}=\frac{1}{4}{\bf B}_{ab}(\tilde{\epsilon}_a\lambda_b
+\tilde{\epsilon}_b\lambda_a)\,,
\end{equation}
and the additional action
\begin{equation}
S_s=-\sum_{a,b}\left[h_{ab}({\bf
E}\phi)_a({\bf
E}\phi)_b\right]\,.
\end{equation}

The actual variation of $S_s$ becomes
\begin{eqnarray}
\delta S_s&=&\sum_c\left[\lambda_c({\bf
E}\phi)_c\sum_a\frac{1}{2}{\bf B}_{ca}({\bf
E}\phi)_a(\tilde{\epsilon})_a\right]\nonumber \\
&-&2\sum_{a,b}\left[h_{ab}({\bf
E}\phi)_a\sum_c{\bf E}_{bc}\epsilon_c\Psi_c\right]\nonumber \\
&=&\sum_c\left[\lambda_c({\bf
E}\phi)_c\sum_a\frac{1}{2}{\bf B}_{ca}({\bf
E}\phi)_a(\tilde{\epsilon})_a\right]\nonumber \\
&-&2\sum_{a,b}\left[h_{ab}({\bf
E}\phi)_a\{({\bf
E}\epsilon)_b\tilde{\Psi}_b+\tilde{\epsilon}_b({\bf
E}\Psi)_b\}\right]\,.
\label{eql}
\end{eqnarray}
In the model of van Nieuwenhuizen~\cite{vN1,vN2,vN3}, the field $h$ is
of course locally coupled to the other fields and the corresponding
term of the last term in Eq.~(\ref{eql}) can be canceled by the
modification of the variation $\delta\lambda_c$. In our case, however,
the term has the contribution of the nearest-neighbor and
next-to-nearest-neighbor variables, so the cancellation needs another
new term.

We have already seen the `non-locality' at the introduction of
$h_{ab}$; thus, it is found to be difficult to obtain the symmetric
model.

The origin of the difficulty is similarly in the fact that the Leibnitz
rule does not hold for difference operators \cite{KSS} in lattice field
theory. Therefore, the `non-locality' grows as the compensating
procedure shown above is advanced.

\subsection{a concise model on a smallest graph}
The approach to finding a `local' supersymmetric model in the previous
subsection has failed for models on a general cycle graph. For a finite
graph, however, the procedure of `supersymmetrization' closes in
finite steps; unfortunately, the continuous limit has no sense of
course in this case. In the present section, we demonstrate the
construction of the model on a smallest path graph, $P_2$.
In this case, the matrix ${\bf E}$ becomes
\begin{equation}
{\bf E}=\left(
\begin{array}{rrr}
0 & 0 & -1\\
0 & 0 & 1\\
1 & -1 & 0
\end{array}
\right)\,,
\end{equation}
which acts on $\phi=(\phi_1~\phi_2~\phi_3)^T$ and
$\Psi=(\Psi_1~\Psi_2~\Psi_3)^T$.
Therefore, the starting action (\ref{2s}) is found to be
\begin{eqnarray}
S_0&=&\frac{1}{2}({\bf
E}\phi)^T{\bf
E}\phi-\frac{1}{2}{\Psi}^T{\bf E}\Psi\nonumber \\
&=&\frac{1}{2}(\phi_1-\phi_2)^2+\phi_3^2-\Psi_3(\Psi_1-\Psi_2)\,.
\end{eqnarray}
At the first time, the super-transformation is defined as follows:
\begin{eqnarray}
& &\delta\phi_1=\epsilon_1\Psi_1\,,\quad
\delta\phi_2=\epsilon_2\Psi_2\,,\quad
\delta\phi_3=\epsilon_3\Psi_3\,,\\
& &\delta\Psi_1=-\phi_3\epsilon_3\,,\quad
\delta\Psi_2=+\phi_3\epsilon_3\,,\quad
\delta\Psi_3=(\phi_1-\phi_2)\frac{1}{2}(\epsilon_1+\epsilon_2)\,,
\label{p2}
\end{eqnarray}
which coincides with the `global' transformation (\ref{gt1}) if
$\epsilon_1=\epsilon_2=\epsilon_3=\epsilon$.

Then the variation $\delta S_0$ turns out to be
\begin{equation}
\delta
S_0=(\epsilon_1-\epsilon_2)(\phi_1-\phi_2)\frac{1}{2}(\Psi_1+\Psi_2).
\end{equation}
This term is expected to be canceled by the variation of the
additional action $S_N$, where
\begin{equation}
S_N=-\lambda(\phi_1-\phi_2)\frac{1}{2}(\Psi_1+\Psi_2)
\end{equation}
with 
\begin{equation}
\delta\lambda=\epsilon_1-\epsilon_2\,.
\end{equation}

Now, the variation of the action $S_0+S_N$ becomes
\begin{eqnarray}
\delta(S_0+S_N)&=&-\lambda(\epsilon_1\Psi_1-\epsilon_2\Psi_2)
\frac{1}{2}(\Psi_1+\Psi_2)-
\lambda(\phi_1-\phi_2)\frac{1}{2}(-\phi_3\epsilon_3+\phi\epsilon_3)
\nonumber
\\ & &=-\lambda\frac{1}{2}(\epsilon_1+\epsilon_2)\Psi_1\Psi_2\,.
\end{eqnarray}
This can be compensated by introducing the following additional
variation of
$\psi_3$:
\begin{equation}
\delta'\Psi_3=-\lambda\frac{1}{2}(\Psi_1\epsilon_1+\Psi_2\epsilon_2)\,.
\end{equation}

To summarize, in this model, the procedure of adding terms to the action
and variations closes at the step $S=S_0+S_N$. Therefore the action
\begin{equation}
S=\frac{1}{2}(\phi_1-\phi_2)^2+\phi_3^2-\Psi_3(\Psi_1-\Psi_2)
-\lambda(\phi_1-\phi_2)\frac{1}{2}(\Psi_1+\Psi_2)\,,
\end{equation}
is invariant under the following fermionic transformation:
\begin{eqnarray}
& &\delta\phi_1=\epsilon_1\Psi_1\,,\quad
\delta\phi_2=\epsilon_2\Psi_2\,,\quad
\delta\phi_3=\epsilon_3\Psi_3\,,\\
& &\delta\Psi_1=-\phi_3\epsilon_3\,,\quad
\delta\Psi_2=+\phi_3\epsilon_3\,,\nonumber \\
& &\delta\Psi_3=(\phi_1-\phi_2)\frac{1}{2}(\epsilon_1+\epsilon_2)
-\lambda\frac{1}{2}(\Psi_1\epsilon_1+\Psi_2\epsilon_2)\,,
\nonumber \\
& &\delta\lambda=\epsilon_1-\epsilon_2\,.
\end{eqnarray}
Note that $S$ is invariant under the exchange
$\phi_1\leftrightarrow\phi_2$ and $\Psi_1\leftrightarrow-\Psi_2$.
Note also that the `superpartner' of the fermionic variable is missing
in this model associated with $P_2$.

A slight modification on the action is incidentally possible and is
shown as
\begin{equation}
S=\frac{1}{2}(\phi_1-\phi_2+\mu\phi_3)^2+\phi_3^2-\Psi_3(\Psi_1-\Psi_2)
-\lambda(\phi_1-\phi_2+\mu\phi_3)\frac{1}{2}(\Psi_1+\Psi_2)\,.
\end{equation}
This action is invariant under the following transformation:
\begin{eqnarray}
& &\delta\phi_1=\epsilon_1\Psi_1\,,\quad
\delta\phi_2=\epsilon_2\Psi_2\,,\quad
\delta\phi_3=\epsilon_3\Psi_3\,,\\
&
&\delta\Psi_1=-\left[\phi_3+\frac{\mu}{2}\left\{
\phi_1-\phi_2+\mu\phi_3-\lambda\frac{1}{2}(\Psi_1+\Psi_2)\right\}\right]
\epsilon_3\,,\nonumber \\
& &\delta\Psi_2=+\left[\phi_3+\frac{\mu}{2}\left\{
\phi_1-\phi_2+\mu\phi_3-\lambda\frac{1}{2}(\Psi_1+\Psi_2)\right\}\right]
\epsilon_3\,,\nonumber \\
&
&\delta\Psi_3=(\phi_1-\phi_2+\mu\phi_3)\frac{1}{2}(\epsilon_1+\epsilon_2)
-\lambda\frac{1}{2}(\Psi_1\epsilon_1+\Psi_2\epsilon_2)\,,
\nonumber \\
& &\delta\lambda=\epsilon_1-\epsilon_2\,.
\end{eqnarray}

\section{Concluding remarks
\label{sec9}}
We have considered various analogue models for supersymmetric theory.
Though they are nothing but toy models, they are in a
class of models which have not ever been focused on.
In higher-derivative analogue models, the simple criterion for
symmetry breaking, i.e., when the partition function vanishes, is yet
not clear.
As far as computation of multiple integral is concerned, 
numerical simulation as lattice field theory may be a key tool to find
the sharp criterion.
The confirmation of the Ward identity in models is also expected
through computational experiments.  

As a simple extension of our models, incorporation of gauge symmetry
can be considered, as in the original dimensional deconstruction.
The non-linear sigma model on a graph is another interesting
model to study.
The analogue models for
Lee-Wick theory with gauge fields and
higher-order Galileon theory are expected to be explored.
The quantum mechanical models provides a playground for 
defining higher-derivative field theories in the model building.

The analysis in the last section implies the difficulty in
constructing `local' models, if models have some continuum limit.
Nevertheless, models on a finite graph seem to be interesting,
because the structure of the models can be interpreted as one of a mass
matrix in a sense of dimensionally-deconstructed field theory.
Thus, the quantum mechanical models would also be an arena of
researching higher-dimensional theory.



\begin{thebibliography}{99}
\bibitem{WittenSQM} E. Witten, Nucl. Phys. {\bf B188} (1981) 513.
\bibitem{BBRT} D. Birmingham, M. Blau, M. Rakowski and G. Thompson,
Phys. Rep. {\bf 209} (1991) 129. 
\bibitem{JMP} N. Kan, K. Kobayashi and K. Shiraishi, J. Math. Phys. 
{\bf 46} (2005) 112301.
\bibitem{PRD} N. Kan, K. Kobayashi and K. Shiraishi, Phys. Rev.
{\bf D80} (2009) 045005.
\bibitem{DD1} N. Arkani-Hamed, A. G. Cohen and H. Georgi, Phys.
Rev. Lett. {\bf 86} (2001) 4757.
\bibitem{DD2} T. Hill,  S. Pokorski and J. Wang, Phys.
Rev. {\bf D64} (2001) 105005.
\bibitem{C1} S. Catterall and E. Gregory, Nucl. Phys. {\bf B487}
(2000) 349.
\bibitem{C2} S. Catterall, JHEP {\bf 05} (2003) 038.
\bibitem{C3} S. Catterall, D. B. Kaplan and M. \"Unsal, Phys. Rep. {\bf
484} (2009) 71. 
\bibitem{K} I. Kanamori, Nucl. Phys. {\bf B841} (2010)
426. 
\bibitem{KS1} T. Kuroki and F. Sugino, Nucl. Phys. {\bf B830} (2010)
434. 
\bibitem{KS2} T. Kuroki and F. Sugino, Nucl. Phys. {\bf B844} (2011)
409. 
\bibitem{KS3} T. Kuroki and F. Sugino, Nucl. Phys. {\bf B867} (2013)
448.
\bibitem{WittenIndex} E. Witten, Nucl. Phys. {\bf B202} (1981) 253.
\bibitem{IJMPA} N. Kan, K. Kobayashi and K. Shiraishi, Int. J. Mod.
Phys.  {\bf A26} (2011) 5369.
\bibitem{GOW} B. Grinstein, D. O'Connell and M. B. Wise, Phys. Rev. {\bf
D77} (2008) 025012.
\bibitem{DGP} G. R. Dvali, G. Gabadadze and M. Porrati,
Phys. Lett. {\bf B485} (2000) 208.
\bibitem{NRT} A. Nicolis, R. Rattazzi and E. Trincherini,
Phys. Rev. {\bf D79} (2009) 064036.
\bibitem{CF} T. Curtright and D. Fairlie,
arXiv:1212.6972 [hep-th].
\bibitem{ADG1} I. Antoniadis, E. Dudas and D.
M. Ghilencea, Nucl. Phys. {\bf B767} (2007) 29.
\bibitem{ADG2} I. Antoniadis, E. Dudas and D. M. Ghilencea, JHEP {\bf
03} (2008) 045.
\bibitem{DPSS} M. Dias, A. Yu. Petrov, C. R. Senise Jr. and A.
J. da Silva, arXiv:1212.5220 [hep-th].
\bibitem{GSS} E. A. Gallegos, C. R. Senise Jr. and A.
J. da Silva, Phys. Rev. {\bf D87} (2013) 085032, arXiv:1212.6613
[hep-th].
\bibitem{KLO} M. K K\"ohn, J.-L. Lehners and B. Ovrut, arXiv:1302.0840.
\bibitem{LW1} 
T. D. Lee and G. C. Wick, 
Nucl. Phys. {\bf B9} (1969) 209.
\bibitem{LW2} 
T. D. Lee and G. C. Wick, 
Phys. Rev. {\bf D2} (1970) 1033.
\bibitem{LW3} 
T. D. Lee and G. C. Wick, 
Phys. Rev. {\bf D3} (1971) 1046.
\bibitem{LW4} 
R. E. Cutkosky, P. V. Landshoff, D. I. Olive and J. C. Polkinghorne, 
Nucl. Phys. {\bf B12} (1969) 281.
\bibitem{LY} Y. Lin, S.-T. Yau,
arXiv:1204.3168 [math.CO].
\bibitem{CLY}
F. Chung, Y. Lin, S.-T. Yau,
arXiv:1207.6612 [math.CO].
\bibitem{BHLLMY}
F. Bauer, P. Horn, Y. Lin, G. Lippner, D. Mangoubi and S.-T. Yau,
arXiv:1306.2561 [math.AP].
\bibitem{vN1} P. van Nieuwenhuizen,
in M. A. Shifman (ed.), ``The many faces of the superworld'',
World Scientific, 2000,
 pp. 649--676.
\bibitem{vN2} 
Contribution to ``Graphs and Patterns in Mathematics and Theoretical
Physics'', Proceedings of Symposia in Pure Mathematics, M.
Lyubich and L. Takhtajan (eds.), 2005,
arXiv:hep-th/0408179.
\bibitem{vN3} P. van Nieuwenhuizen and D. V. Vassilevich,
Class. Quant. Grav. {\bf 22} (2005) 5029-5051,
hep-th/0507172.
\bibitem{KSS}
M. Kato, M. Sakamoto and H. So,
arXiv:hep-lat/0509149 [hep-lat],
JHEP {\bf 05} (2008) 057,
(arXiv:0803.3121 [hep-lat]),
arXiv:0810.2360 [hep-lat],
arXiv:1212.1533 [hep-lat].
\end{thebibliography}
\end{document}